\documentclass[useAMS, usenatbib]{mn2e}
\usepackage[dvips]{graphicx}
\usepackage{amssymb}
\usepackage{multirow}

\input{epsf}

\newcommand{\kms}{\thinspace km\,s$^{-1}$}
\newcommand{\msol}{${\mathrm M_{\odot}\thinspace}$}

\newcommand{\ha}{\mbox{H$\alpha$}}

\newcommand{\kmsmpc}{\thinspace km\thinspace s$^{-1}$\thinspace Mpc$^{-1}$}

\title[Dynamical virial masses of LBGs]{Dynamical virial masses of
  Lyman--break galaxy haloes at {\boldmath $z=3$}} \author[S. J. Weatherley \&
  S. J. Warren] {Stephen~J.~Weatherley\thanks{Email :
  stephen.weatherley@imperial.ac.uk} and Stephen~J.~Warren\\
  Astrophysics Group, Blackett Laboratory, Imperial College London,
  Prince Consort Road, London SW7 2BW, UK}
	
\date{Accepted 0000 January 00.
      Received 0000 January 00;
      in original form 0000 January 00}

\pagerange{\pageref{firstpage}--\pageref{lastpage}}
\pubyear{2005}
\begin{document}
\maketitle
\label{firstpage}


\begin{abstract}

We improve on our earlier dynamical estimate of the virial masses of
the haloes of Lyman-break galaxies (LBGs) at redshift $z=3$ by
accounting for the effects of seeing, slit width, and observational
uncertainties. From an analysis of the small number of available
rotation curves for LBGs we determine a relation
$V_{c7}=(1.9\pm0.2)\sigma$ between circular velocity at a radius of
7\thinspace kpc, $V_{c7}$, and central line velocity width,
$\sigma$. We use this relation to transform the measured velocity
widths of 32 LBGs to the distribution of circular velocities,
$V_{c7}$, for the population of LBGs brighter than ${\mathcal
R=25.5}$. We compare this distribution against the predicted
distribution for the `massive--halo' model in which LBGs pinpoint all
of the highest mass dark matter haloes at that epoch. The observed LBG
circular velocities are smaller than the predicted circular velocities
by a factor $>1.4\pm0.15$. This is a lower limit as we have ignored
any increase of circular velocity caused by baryonic dissipation.  The
massive--halo model predicts a median halo virial mass of
$10^{12.3}$\msol , and a small spread of circular velocities,
$V_{c7}$. Our median estimated dynamical mass is $<
10^{11.6\pm0.3}$\msol , which is significantly smaller; furthermore,
the spread of our derived circular velocities is much larger than the
massive--halo prediction.  These results are consistent with a picture
which leaves some of the most--massive haloes available for occupation
by other populations which do not meet the LBG selection criteria. Our
new dynamical mass limit is a factor three larger than our earlier
estimate which neglected the effects of seeing and slit width. The
median halo mass recently estimated by Adelberger et al. from the
measured clustering of LBGs is $10^{11.86\pm0.3}$\msol . Our dynamical
analysis appears to favour lower masses and to be more in line with
the median mass predicted by the collisional starburst model of
Somerville et al., which is $10^{11.3}$\msol.
\end{abstract}


\begin{keywords}

galaxies: kinematics and dynamics -- galaxies: formation -- galaxies:
high redshift

\end{keywords}


\section{Introduction}

The development of the Lyman--break selection technique (Steidel \&
Hamilton, 1993) led to the first substantial samples of galaxies at
high redshifts. There is now a wealth of observational data
on this population, including optical and near-infrared photometry;
optical and near-infrared spectroscopy; and measurement of their
clustering properties (Steidel, Pettini \& Hamilton, 1995; Shapley
et al., 2001; Shapley et al., 2003; Pettini et al., 2001; Adelberger et al.,
1998; Giavalisco \& Dickinson, 2001). Nonetheless, the most
fundamental quantity -- the dark--matter halo virial mass, or
equivalently the circular velocity at large radius -- is poorly
determined for Lyman--break galaxies (LBGs). This is because it requires
observations in the very faint outer regions of galaxies of rest--frame optical emission lines, 
which are redshifted to the near--infrared where the night sky is bright.
As a consequence only a
handful of rotation curves, with low signal--to--noise and compromised by seeing,
have been measured for LBGs. This paper is concerned with extracting
the maximum useful information on the virial masses of the haloes of
LBGs from the existing kinematic data. Throughout the paper the term LBG
refers to galaxies at $z=3$ identified by their spectral discontinuity
at the Lyman--limit, that are brighter than ${\mathcal R}=25.5$ (Adelberger
et al., 1998). 

Lacking high quality rotation curves, most of our knowledge of the
virial masses of LBGs comes indirectly, from an interpretation of their
clustering properties. The strong clustering measured in the
counts--in--cells analysis of Adelberger et al. (1998) and in the
correlation--function analysis of Giavalisco \& Dickinson (2001), 
and the dependence of the correlation length on luminosity detected in the
latter study, were interpreted as implying that LBGs are associated
with dark matter haloes of high mass -- consistent with a simple model in
which LBGs pinpoint all the highest mass dark matter haloes at that
epoch. We refer to this model  as the `massive--halo' model and
use the results of the analysis by Mo, Mao \& White (1999; hereafter
MMW) as defining the model. Their figure 4 provides the predicted
distribution of halo circular velocities at the virial radius, and
yields a median halo virial mass of $10^{12.3}$\msol for the LBG
population. The massive--halo model leaves no room for other
high--redshift populations, that do not meet
the LBG criteria, to occupy the high--mass tail of the halo
distribution, such as sub-mm galaxies (Smail, Ivison \&
Blain, 1997) and red galaxies (Franx et al., 2003).
Adelberger et al. (2005) have recently measured the
clustering of a new, larger, sample of LBGs: through comparison with
n--body simulations they estimate a median halo mass of
$10^{11.86\pm0.3}$\msol. The best--fit value for the median mass is a
factor three lower than in the massive--halo model, in which case not all of
the most--massive haloes would be occupied by LBGs -- leaving
some space for other galaxy types.

Unfortunately the clustering results allow for more than one
interpretation. As shown by Wechsler et al. (2001) an alternative
model proposed by Somerville, Primack \& Faber (2001; hereafter SPF),
in which LBGs have lower masses but are temporarily brightened due to
merger--induced star formation, predicts similarly strong
clustering as the massive--halo model. In this picture the virial
masses are an order of magnitude lower than in the massive--halo model,
with median value of $10^{11.3}$\msol (Primack, Wechsler \&
Somerville, 2003).

Finally, an indirect lower limit to the halo masses is provided by
multiplying the estimated stellar masses of LBGs by the ratio of the
cosmological matter density to the density in baryons
($\Omega_m/\Omega_b$). Taking the median value of the stellar mass
estimated by Shapley et al. (2001), from fits to optical and
near-infrared photometry, this provides a lower limit to the median
halo virial mass of $10^{11.1}$\msol.

To summarise, current estimates of the median halo virial mass of LBGs
at redshift $z=3$, which are all indirect, cover a wide range, $10^{11.1}$\msol
to $10^{12.3}$\msol. Motivated by this large uncertainty, in an
earlier paper (Weatherley \& Warren, 2003; hereafter Paper I) we
presented a simplified analysis of the kinematic data on LBGs, to
estimate the halo virial masses. The $\Lambda$CDM paradigm for 
structure formation permits computation of the detailed properties
of the dark matter yielding predictions of the masses of haloes, their
mass profiles, and their assembly history (e.g. Lacey \& Cole,
1993; Navarro, Frenk \& White, 1997, hereafter NFW). Baryonic
processes, including star formation and feedback, on the other hand
are too complex to allow {\em ab initio} predictions of the
distribution and state of the baryons within the haloes. Therefore the best
tests of the $\Lambda$CDM paradigm will be those which measure the
properties of the dark matter. Instead, at high redshift, we are
largely limited to studies of luminous matter in the brightest central
regions of galaxies, where the baryons dominate the mass budget. For
example, in most cases the kinematic data on LBGs is limited to the
measured line velocity widths. These data, on their own, are not useful for
measuring halo masses; fortunately, in a handful of cases spatially
resolved data have been obtained, extending to projected radii of a
few kpc.

Our dynamical analysis centred on these rotation velocities, $V_r$. In
Paper I we collated the useful kinematic data on LBGs, selecting from
the datasets of Pettini et al. (2001) and Erb et al. (2003), to define
a sample of seven measured projected rotation velocities, $V_r$, as
well as 32 velocity widths, characterised by the dispersion of the
velocity profile $\sigma$.  The galaxies for which rotation curves
have been measured may not be representative of the LBG population:
for example the luminosity profiles may be more extended than
average. For this reason, a simple statistical analysis of the
rotation velocities is not appropriate. Instead we were able to use
the rotation velocities by making the assumption that the distribution
of measured values of the dimensionless quantity $V_r/\sigma$ are
representative. We then connected the rotation velocities, the
velocity widths, and the $\Lambda$CDM predictions through the
assumption of a linear relation $V_{c7}=\alpha\sigma$, where $V_{c7}$
is the circular velocity at a radius of 7kpc.\footnote{Our notation is
as follows: rotation velocity, $V_r$, refers to half the measured
peak--to--peak velocity spread of a spatially resolved velocity
profile (the observed quantity).  Circular velocity, $V_c$, refers to
the de--projected, true, circular velocity of the galaxy. Similarly,
we use the symbol $r$ for projected radius and $R$ for de--projected
radius.}  The procedure was, firstly, to determine an observed value
of $\alpha(obs)$ which, accounting for random orientations and
inclinations (over cos($i$)), explained the observed distribution of
$V_r/\sigma$ of the seven galaxies. We found that a value
$\alpha(obs)=\sqrt{2}$ (motivated by the isothermal sphere model)
provided a satisfactory fit.  This value of $\alpha(obs)$ then
converts the distribution of the 32 values of $\sigma$, assumed
representative of the population, to the distribution of $V_{c7}$ for
LBGs. We compared this against the, larger, required value of
$\alpha(req)=2.6$ that would be needed to reconcile the measured
distribution of $\sigma$ with the distribution of $V_{c7}$ predicted
by the massive--halo model. The small value of $\alpha(obs)$ was
inconsistent with the required value of $\alpha(req)$, implying halo
virial masses an order of magnitude smaller than the predictions of
the massive--halo model.

Our previous analysis made a number of simplifying assumptions. In
computing $\alpha(obs)$ we took no account of the effect of
atmospheric seeing, which would blur the light from the bright, baryon
dominated centres of the galaxies to the faint outer parts, smoothing
the transition from $+$ve to $-$ve velocity in the rotation curve,
such that the rotation velocity would be underestimated. The data we
analysed were taken in average conditions of 0.5\thinspace arcsec seeing, and
the rotation velocities were measured at an average offset of
0.64\thinspace arcsec. Erb et al. (2004) give an indication of the importance
of seeing: in 0.9\thinspace arcsec seeing they measured a rotation velocity
approximately half that measured for the same LBG in 0.5\thinspace arcsec
seeing.  As discussed later, and illustrated in
Figure \ref{fig:lbg_rotationcurve}, we find that the reduction in
measured rotation velocity in 0.5\thinspace arcsec seeing is indeed
significant. Another
important effect not accounted for in Paper I is the variation of
projected velocity across the finite width of the slit (0.76\thinspace arcsec
or 1.00\thinspace arcsec). The influences of seeing; slit width; galaxy
luminosity profile; galaxy orientation and inclination; the
observational errors; and selection cuts all interact in a complex
manner. In this paper we improve on our previous study by
detailed modelling of these various effects and we investigate the
sensitivity of our results to the assumptions made. We retain the
formalism of the relation $V_{c7}=\alpha\sigma$ as the means of
comparing theory and data. We employ a likelihood analysis, improving
over the statistical analysis of Paper I.
Throughout, we assume a standard, flat $\Lambda$CDM cosmology with
$\Omega_\Lambda = 0.7$ and H$_0 = 70$\kmsmpc.


\section{Revised analysis}

We follow the same principles as in our earlier analysis. The data
analysed are the same seven rotation velocities, $V_r$, and 32 velocity widths, $\sigma$,
 summarised in Paper I. Below, we compare
the $\sigma$ distribution for the LBGs against the predicted
distribution of $V_{c7}$ for the massive--halo model to determine
$\alpha(req)$. Secondly we reanalyse the seven measured rotation velocities
and their associated values of $\sigma$, modelling the observational effects
and selection cuts, as detailed below, to measure $\alpha(obs)$ --
i.e. the value of $\alpha$ that gives the best fit to the distribution
of $V_r/\sigma$ for the seven galaxies. We then compute the probability 
of observing the seven values of $V_r/\sigma$ assuming 
$\alpha$=$\alpha(req)$ --
 i.e. the probability that the
massive--halo model fits the data. 

At this point it is worth repeating,
as emphasised in Paper I, that we make no assumption about the cause
of the line broadening i.e. the relative contribution of pressure and
rotational support to the velocity width. Our analysis is merely aimed at
testing whether the observed seven values of the ratio $V_r/\sigma$ are
large enough to reconcile the $\sigma$ distribution for LBGs with the
massive--halo model.

\subsection{Determining {\boldmath $\alpha(req)$}}

\begin{figure}
\begin{center}
\includegraphics[width=0.85\columnwidth]{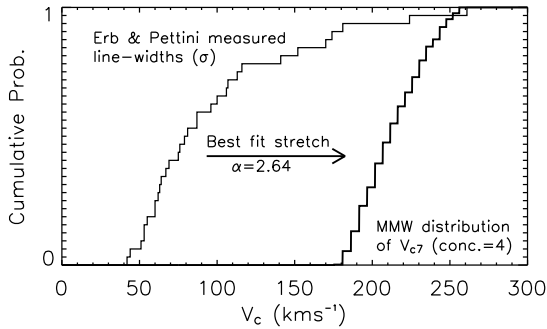}
\caption{\label{fig:lbg_mmwdistribution}The cumulative PDF of the
line velocity widths (Paper I, table 1) for LBGs compared with the
distribution of $V_{c \rm 7}$ predicted by the massive--halo model. A
value $\alpha(req)=2.64$ is required to reconcile the two curves.}
\end{center}
\end{figure}

Figure \ref{fig:lbg_mmwdistribution} plots the cumulative PDF of the
velocity widths (corrected for instrumental resolution) 
of
the 32 LBGs (thick stepped line), taken from table 1, Paper I. For the
seven galaxies with spatially resolved velocity profiles, $\sigma$ was
measured from the central line profile.

Also plotted in Figure \ref{fig:lbg_mmwdistribution} is the MMW
prediction for the distribution of halo circular velocities at a
radius of 7\thinspace kpc (thin stepped line). This was calculated
starting with the distribution
of halo velocities at the virial radius $V_h$ from figure 4 of MMW. For
the NFW profile, the conversion to circular velocity at radius R is
\[ 
\Big(\frac{V_{c \rm R}}{V_h}\Big)^2=\frac{1}{x}\frac{\ln(1+cx)-(cx)/(1+cx)}
{\ln(1+c)-c/(1+c)},
\]
where $x=R/R_h$.  We use the relation $R_h=V_h/(10H(z))$ and we assume
concentration $c=4$. These are the circular velocities computed
ignoring the effects of baryonic dissipation, which will raise
the circular velocities. This increase is probably substantial but is
difficult to predict reliably. The simple adiabatic contraction model (Mo, Mao
\& White 1998) indicates that the circular velocities $V_{c7}$ could increase
by a factor of about 1.5, but the detailed processes involved in baryonic
contraction are poorly understood. There is both theoretical
(El-Zant, Shlosman, \& Hoffman 2001) and observational (Binney \&
Evans 2001) evidence to suggest that the increase may not be as great.

A stretch of the Erb and Pettini velocity widths of $\alpha(req)=2.64$ gives the best fit to the predicted 
distribution of $V_{c7}$, as shown in Figure 1. It is noticeable that the predicted distribution of $V_{c7}$ is rather
narrow, compared to the broad distribution of $\sigma$. We consider
this point again later.

\subsection{Modelling observational effects and selection cuts to
  determine {\boldmath $\alpha(obs)$}} 

\begin{figure}
\begin{center}
\includegraphics[angle=90, width=0.85\columnwidth]{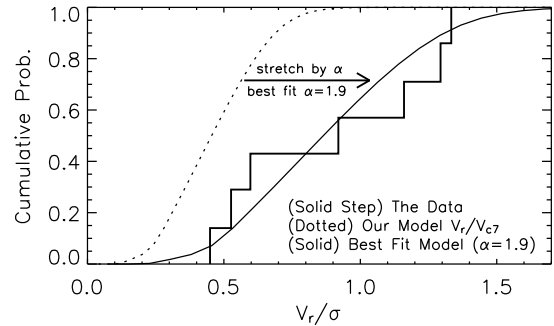}
\caption{\label{fig:lbg_cumf} Cumulative PDF of the ratio $V_r/\sigma$
for the seven galaxies of Pettini et al. and Erb et al. (table 2 in
Paper I) shown as the solid stepped line.  The dotted line shows the
ratio of $V_r/V_{c\rm 7}$ for our model galaxies, including errors and
selection criteria.  The solid smooth line shows the same curve,
stretched by the best fit value of $\alpha=\alpha(obs)$, to match our
model to the data.}
\end{center}
\end{figure}

The principle of measuring $\alpha(obs)$, accounting for observational
effects and selection cuts, is illustrated in Figure \ref{fig:lbg_cumf},
and is explained as follows. 
For galaxies of specified circular
velocity radial profile, and luminosity radial profile, it is possible
to predict the observed distribution of the ratio
$V_r/V_{c7}$, for any given
observing conditions -- defined by the seeing, the slit width
and the selection criteria.
Since $\alpha=V_{c7}/\sigma$ we can stretch our model distribution of $V_r/V_{c7}$
by a factor $\alpha$ to match the observed distribution of
$V_r/\sigma$. This best fit stretch factor is $\alpha(obs)$. If the massive--halo
model is correct $\alpha(obs)$ will be consistent with
$\alpha(req)$. As regards the slit width, since two observational
set--ups were used, with 0.76\thinspace arcsec (NIRSPEC) and 1.00\thinspace arcsec
(ISAAC) slits, we model both cases, and weight the results by the
fraction of observations with each slit. We now explain the modelling
in detail.

The galaxy rotation curves are assumed to reflect the halo mass
profile; therefore the galaxies are modelled as massless stellar discs
of exponential luminosity profile -- the choice of scale radius, $R_0$, is
discussed below.  The distribution of galaxy circular velocity 
profiles is defined by the MMW halo circular velocity distribution,
and the NFW mass profile, with concentration $c=4$. No information on
the inclination angles and position angles of the major axes were
available at the time the spectra were acquired; therefore, we assume
random inclination angles (over $\cos(i)$), and random orientations relative to the
slit. The 2D luminosity--tagged velocity maps are convolved with the
seeing, and the resulting luminosity--weighted velocity maps are
averaged across the slit to produce luminosity--weighted observed
rotation curves along the slit.

\begin{figure}
\begin{center}
\includegraphics[angle=90, width=0.85\columnwidth]{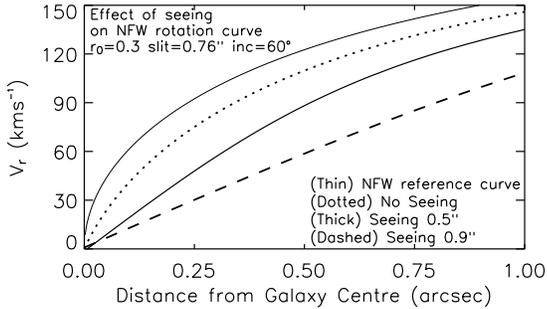}
\caption{\label{fig:lbg_rotationcurve}The effect of seeing on the
measured rotation curve of a galaxy with halo velocity $V_h=220$\kms\
inclined at 60 degrees to the sky, with major--axis aligned with the spectroscopic slit.}
\end{center}
\end{figure}

Example rotation curves for a galaxy
with $V_h=220$\kms, and inclination angle $i=60^{\circ}$, oriented
along the slit, are provided in Figure \ref{fig:lbg_rotationcurve}. The
curves from top to bottom represent: i) the NFW circular velocity
profile along the major axis; ii) the luminosity-weighted rotation
curve obtained by averaging across the slit, before convolution by the
seeing; iii) the observed rotation curve in 0.5\thinspace arcsec seeing; iv)
the observed rotation curve in 0.9\thinspace arcsec seeing. At a position
0.75\thinspace arcsec along the slit, the measured rotation velocities are
0.85 and 0.62 of the NFW value on the major axis, for 0.5\thinspace arcsec and
0.9\thinspace arcsec seeing respectively.

To compare the model with the data we must reproduce the conditions by
which the data were measured, as closely as possible. For a chosen
scale radius $R_0$, we measured the rotation velocity at the
(deprojected) radius $\beta R_0$, where the value of the constant
$\beta$ was chosen to reproduce the average projected radius of the
measured rotation curves.  For the rotation curve to be spatially
resolved the projection of this distance on the slit had to be greater
than some limiting value, which we took as $r_{lim}=0.4$\thinspace arcsec.  If
the simulation passes the spatial cut we add a velocity error drawn
at random from an appropriate Gaussian distribution and then check if
the observation meets the velocity cut, $V_{min}$. Rotation velocities were only
recorded by Pettini et al. (2001) and Erb et al. (2003) if a velocity
gradient was clearly visible, corresponding to a rotation velocity
greater than twice the typical velocity error (Erb, private
communication). Therefore, in our simulation, we disregard rotation velocities
smaller than $V_{min}=25$\kms\ for the ISAAC data, and
$V_{min}=50$\kms\ for the NIRSPEC data. We chose a nominal scale
length $R_0=0.3$\thinspace arcsec, and found that a value $\beta=2.6$ reproduced
the average projected radius of the seven measurements. This scale radius
is about twice as large as the typical value measured for LBGs
(Marleau \& Simard, 1998); nevertheless, given that rotation
velocities were secured for only seven of the 32 LBGs considered here, it
may be true that the light profiles of these seven galaxies are more
extended on average.

In summary, our model is fully specified by the distribution of halo
virial circular velocities from MMW and the following list of parameters (fiducial 
values in brackets):

\begin{enumerate}

\item{{\boldmath $\sigma_{\rm atm}$} --- the seeing with which we
convolve the galaxy profiles (FWHM=0.5\thinspace arcsec, the quoted average
value from Pettini et al. (2001), and Erb et al. (2003))}

\item{{\boldmath $c$} --- the concentration parameter of the NFW 
halo profile (4)}

\item{{\boldmath $R_0$} --- scale--radius for the exponential light
profile (0.3\thinspace arcsec); and the spatial cut {\boldmath $r_{lim}$}
(0.4\thinspace arcsec)}

\item{{\boldmath $V_{min}$} --- the spectral cut (25\kms\ for the
  ISAAC 1.00\thinspace arcsec slit and 50\kms\ for the NIRSPEC 0.76\thinspace arcsec
  slit)}

\end{enumerate}

Later we assess how robust are the conclusions to variations in the
chosen values of the above parameters.


\section{Results}

\subsection{Measuring {\boldmath $\alpha(obs)$}}

The results of the modelling are summarised in the curve of the
expected distribution of $V_r/V_{c7}$ plotted in
Figure \ref{fig:lbg_cumf}. Also plotted in the figure, as the stepped
line, is the distribution of $V_r/\sigma$ for the seven galaxies with
recorded rotation velocities. Recall that $\alpha(obs)$ is the value
by which the curve should be stretched to fit the stepped line, and
that this should be consistent with $\alpha(req)=2.64$
(Figure \ref{fig:lbg_mmwdistribution}) if the massive--halo model
is correct.

\begin{figure}
\begin{center}
\includegraphics[angle=90, width=0.85\columnwidth]{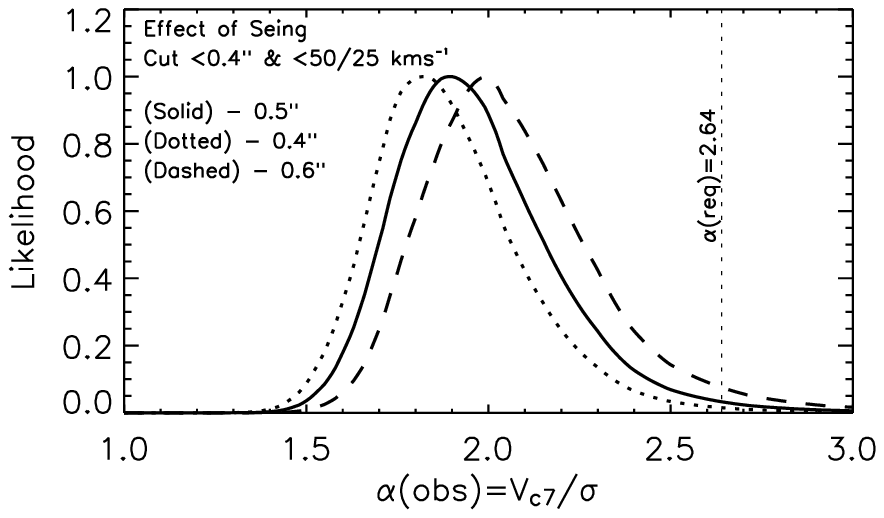}
\includegraphics[angle=90, width=0.85\columnwidth]{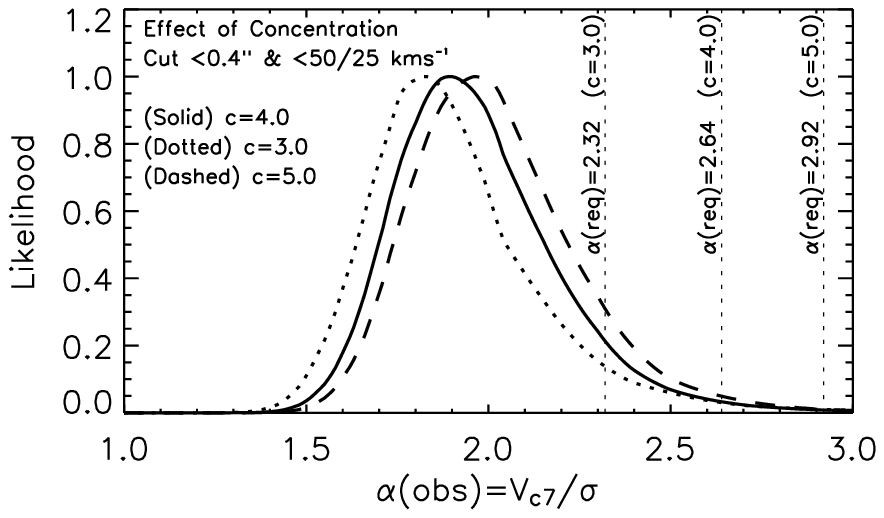}
\caption{\label{fig:lbg_bayesians}Likelihood plots (normalised to the
peak) to show the effects of changing parameters in the
model. The upper plot shows the effect of changing the atmospheric
seeing, and the lower plot shows the effect of varying the concentration.
All other parameters are kept fixed at the preferred values.}
\end{center}
\end{figure}

In Figure \ref{fig:lbg_bayesians}(upper) we plot, as the solid curve,
the likelihood of the data as a function of $\alpha$, normalised to
the peak. This is the posterior pdf (modulo a constant) for
$\alpha$ adopting a uniform prior. The best fit, plotted as the
stretched curve in Figure \ref{fig:lbg_cumf} is $\alpha(obs)=1.9\pm0.2$.
The probability of the data if $\alpha$ is as large or larger than
$\alpha(req)=2.64$ is $p=0.013$. This means that the measured rotation
velocities are too small to be consistent with the massive--halo
model. Another way of visualising this result is to say that the
rotation velocities imply a distribution of $V_{c7}$ for LBGs which is
the distribution of $\sigma$ plotted in
Figure \ref{fig:lbg_mmwdistribution} stretched by the factor
$\alpha(obs)=1.9$. The circular velocities are smaller than the
predicted circular velocities for the massive--halo model by a factor
$1.4\pm0.15$.

\subsection{Variation of model parameters}\label{sec:lbg_effect}

In this section we explore to what extent variation of the parameters
of the model affects the conclusion that LBGs are significantly less
massive than predicted by the massive--halo model.

\begin{enumerate}

\item We find that our results are not very sensitive to the exactly
  value of the seeing adopted. We repeated the entire analysis for
  values of 0.4\thinspace arcsec and 0.6\thinspace arcsec. The corresponding
  likelihood curves are shown in the top plot of
  Figure \ref{fig:lbg_bayesians}, and compared to the curve for the
  fiducial value of 0.5\thinspace arcsec. Over the range
  0.4 -- 0.6\thinspace arcsec the probability for the massive--halo model
  varies over the range $0.006<p<0.032$.

\item The lower plot shows the effect of varying the value of the
  concentration parameter $c$. Since the relation between $V_{c \rm
  7}$ and $V_h$ depends on $c$, varying $c$ changes not only the
  predicted rotation velocities but also the value of $\alpha(req)$,
  as indicated on the plot.  For the range of $V_h$ specified by MMW 
  a suitable range for the concentration parameter is $3<c<5$ (Bullock et al. 2001).
  For this range, the probability for the massive--halo model varies over the range
  $0.059>p>0.004$. Therefore, for a spread of values of $c$ within the quoted
  range, the probability for the massive--halo model remains low.

\item The results are rather insensitive to small variations of the
  exponential scale radius, $R_0$, and the spatial cut $r_{lim}$. This
  is because for variations in either parameter we have to adjust the
  parameter $\beta$ such that the average projected radius at which
  the rotation velocity is measured matches that of the data. We
  confirmed that for smaller $R_0$ the measured $\alpha(obs)$ is
  slightly higher, and {\em vice versa}. This is as expected because
  for smaller $R_0$ the light is more concentrated, so after
  convolution with the seeing the luminosity--weighted rotation
  velocity at any radius receives proportionally more light
  originating at smaller radius, where the rotation velocity is lower.

\item We also checked the sensitivity of the result to the value of
  the spectral cut $V_{min}$. Our analysis uses the spectral cuts
  applied to the data $V_{min}=50$\kms\ and $V_{min}=25$\kms\ for the
  NIRSPEC and ISAAC datasets respectively, and we have taken these at
  face value. Nevertheless it is interesting to check how the
  conclusions would have differed, had different cuts been
  applied. Higher cuts make little difference, but lower cuts raise
  the likelihood of high values of $\alpha$.  For example had the same
  distribution of $V_r/\sigma$ been observed for cuts as low as low as
  $V_{min}=30$\kms\ and $V_{min}=15$\kms , for the NIRSPEC and ISAAC
  datasets, the data would be consistent with the massive-halo model,
  with $p=0.15$. This may be understood by reference to Figure
  \ref{fig:lbg_cumf}. Higher values of $V_{min}$ raise the low--end
  cut off of the predicted curve of $V_r/\sigma$. High values of
  $\alpha$ are ruled out once the tail passes the fist point on the
  stepped line. Lowering the cuts extends the tail of the curve to
  smaller values of $V_r/\sigma$, so that larger values of $\alpha$
  are permitted. This illustrates the importance of specifying
  well--defined selection criteria.

\end{enumerate}

\section{Discussion} \label{sec:lbg_discussion}

To summarise the previous section, we have improved on our earlier
dynamical estimate of the virial masses of the haloes of LBGs at
redshift $z=3$, by accounting for the effects of seeing, slit width,
and observational uncertainties. From an analysis of the small number
of available rotation curves for LBGs we determined a relation
$V_{c7}=(1.9\pm0.2)\sigma$. We used this to transform the measured
velocity widths of 32 LBGs to the distribution of circular velocities
$V_{c7}$. We compared this distribution against the predicted
distribution for $V_{c7}$ for the massive--halo model.  The LBG
circular velocities are too small by a factor $1.4\pm0.15$.  As noted
earlier, our analysis does not account for the increase in the
circular velocity of the galaxies due to baryonic dissipation, so the
quoted factor is a lower limit. With this in mind we can compute an upper
limit to the median dynamical halo virial mass, by taking the median
value of $\sigma$, and multiplying by 1.9 to convert to
$V_{c7}$. Selecting an appropriate value of $c$, our median estimated
dynamical halo mass is $<10^{11.6\pm0.3}$\msol. Our new dynamical mass
limit is a factor 3 larger than our earlier estimate (Paper I) which
neglected the effects of seeing and slit width. Our dynamical mass
limit is inconsistent with the value of $10^{12.3}$\msol predicted by
the massive--halo model. The broad spread in measured velocity widths
(Figure \ref{fig:lbg_mmwdistribution}), which translates to a broad
spread in inferred $V_{c7}$ is further evidence against the
massive--halo model, which predicts a narrow spread in $V_{c7}$. A
consistent picture is one in which LBGs inhabit only a fraction of the
most--massive haloes. In this case the average mass will be lower and
the range of masses will be larger.  This conclusion finds some
support in the recent clustering analysis of Adelberger et al. (2005):
by matching the clustering strength, and space density to the
corresponding quantities for haloes in an n--body simulation they
derive a median halo mass of $10^{11.86\pm0.3}$\msol. Our analysis
suggests lower masses still.  Our result appears to be more consistent
with the predicted masses of the collisional starburst model of SPF,
which has a median halo mass of $10^{11.3}$\msol. The picture that
emerges is one which leaves some of the most--massive haloes available
for occupation by other populations which do not meet the LBG
selection criteria.

These conclusions rest on the analysis of a small number of rotation
curves of low signal--to--noise, without the benefit of complementary
high--resolution imaging. Recently Erb et al. (2004) have questioned
whether the majority of LBGs are dynamically relaxed, which would undermine the
basis of our calculation. In an attempt to maximise the chances of
measuring rotation, they used HST imaging of a sample of galaxies at
$z\sim 2$ to select the fraction with morphologies consistent with the
expectation for edge-on disks. They obtained \ha\ spectra for nine
of their sample, with the slit aligned along the major axis. Contrary
to expectation, rotation velocities were secured for only two of the
galaxies. They proposed that these results could be explained if some
of the elongated galaxies are in fact merging sub-units, rather than
relaxed edge--on disks. It is worth noting, nevertheless, that the
majority of these observations were undertaken in typical seeing of
0.8\thinspace arcsec, somewhat worse than for the data analysed here. Of the nine
galaxies selected as elongated, in six cases the (deconvolved) extent of
the \ha\ emission is smaller than the seeing, which would make it
difficult to detect rotation. Of the three cases where the \ha\
emission is more extended than the seeing, rotation velocities were
secured for two. This analysis further underscores the desirability of
high--spatial resolution 2D spectroscopy, to remove some of these
uncertainties of interpretation. In the meantime, our goal in this
paper has been to maximise the useful information on the dynamical
virial masses of the haloes of LBGs from existing kinematic data.

\section*{Acknowledgments}

We thank Dawn Erb and Max Pettini for their helpful comments regarding
their data, and their
constructive criticism of Paper I. We are grateful to the referee for comments
which helped clarify this manuscript. SJW[1] also thanks Thomas Babbedge
for his comments on previous drafts of this manuscript.


\bsp

\label{lastpage}

\end{document}